\documentstyle[11pt,aasms4,rotating]{article}

\def\simlt{\lower.5ex\hbox{$\; \buildrel < \over \sim \;$}}
\def\simgt{\lower.5ex\hbox{$\; \buildrel > \over \sim \;$}}

\def\gcm3{{\rm\,g\,cm^{-3}}}
\def\ncm3{{\rm\,cm^{-3}}}

\def\>{$>$}
\def\<{$<$}

\def\refindent{\par\noindent\hangindent=3pc\hangafter=1 }

\def\apj#1#2#3{\refindent#1, {\it ApJ}, {\bf#2}, #3.}

\def\mnras#1#2#3{\refindent#1, {\it MNRAS}, {\bf#2}, #3.}

\lefthead{Melia et al.}
\righthead{}

\begin{document}
\centerline{Submitted to the Editor of the Astrophysical Journal Letters}
\vskip 0.5in
\title{\bf Polarimetric Imaging of the Massive Black Hole at the Galactic Center}

\author{Benjamin C. Bromley,$^{1}$ Fulvio Melia,$^{2,3,4}$ and
Siming Liu$^{2}$}

\affil{$^1$Department of Physics, University of Utah, 201 JFB, Salt Lake City, 
UT 84112}
\affil{$^2$Physics Department, The University of Arizona, Tucson, AZ 85721}
\affil{$^3$Steward Observatory, The University of Arizona, Tucson, AZ 85721}


\altaffiltext{4}{Sir Thomas Lyle Fellow and Miegunyah Fellow.}


\begin{abstract}

The radio source Sgr A* in the Galactic center emits a
polarized spectrum at millimeter and sub-millimeter wavelengths that
is strongly suggestive of relativistic disk accretion onto a massive
black hole. We use the well-constrained mass of Sgr A* and a
magnetohydrodynamic model of the accretion flow to match both the
total flux and polarization from this object. Our results demonstrate
explicitly that the shift in the position angle of the polarization vector, 
seen at wavelengths near the peak of the mm to sub-mm emission from this
source, is a signal of relativistic accretion flow in a strong 
gravitational field.  We provide maps of the polarized emission to 
illustrate how the images of polarized intensity from the vicinity
of the black hole would appear in upcoming observations with very 
long baseline radio interferometers (VLBI). Our results suggest that 
near-term VLBI observations will be able to directly image the polarized
Keplerian portion of the flow near the horizon of the black hole.

\end{abstract}




%

\section{Introduction}
The compact radio source Sgr A* at the Galactic center continues to
draw attention from both theorists and observers as the growing body
of evidence points more and more to a massive black hole paradigm to
explain its characteristics and behavior (for the latest comprehensive
review on this subject, see Melia \& Falcke 2001).  The kinematics of
the central star cluster probes the gravitational potential within a
mere $0.015$ pc of the nucleus and reveals the presence of $2.6\times
10^6\,M_\odot$ of dark matter in this region (Genzel et al. 1997; Ghez
et al. 1998).  Several alternative scenarios invoked to explain this
condensation of matter have been excluded with a high degree of confidence 
(Haller et al. 1996; Genzel et al. 1996; Melia \& Coker 1999).

Complementary information on the nature of Sgr A* is provided by its spectral 
properties, which reflect the physics of matter accreting toward Sgr A* at 
distances of several Schwarzschild radii ($r_S\equiv 2GM/c^2\approx 7.7\times 
10^{11}$ cm, or roughly $0.05$ A.U.).  Falcke, Melia \& Agol (2000) have shown 
that the `shadow' cast by the black hole may have an apparent diameter of about
$5\,r_S$, nearly independent of the black hole spin or its orientation.
This depression in intensity (literally, a `black hole') arises from the
effects of strong light bending near the event horizon, so that radiation
emitted in the background is either absorbed or redirected away from the
line-of-sight.  The distance to Sgr A* being roughly $8$ kpc, this diameter
corresponds to an angular size of $\sim 30\,\mu$as, which approaches the
resolution of current radio interferometers.

This is very exciting because several aspects of the observational program 
are evolving toward a fortuitous convergence to make such an important 
observation feasible within just a few years.  No other galactic nucleus,
whether in the local and M81 groups, or in the category of very active 
(though much more distant) sources, can provide such an opportunity in the
foreseeable future.  Sgr A*'s spectrum at mm/sub-mm wavelengths shows a
distinctive `hump' (Falcke et al. 1998), whose characteristics seem to
confirm the suggestion that the radiating plasma (emitting predominantly
via thermal synchrotron processes) is making a transition from optically 
thick to thin emission near $\nu=2\times 10^{11}$ Hz (Melia 1992, 1994).
Other scenarios, such as the jet model (Falcke \& Markoff 2000) and
the ADAF disk model (Narayan et al 1995), also require that the medium
goes transparent across this range of frequencies.  The fortunate aspect
about this transition is that it happens to occur at about the same
set of frequencies where the scatter broadening in the interstellar
medium decreases below the intrinsic source size (see also Melia,
Jokipii \& Narayanan 1992).  Falcke, Melia \& Agol (2000) showed that 
taking this effect into account, as well as the finite achievable 
telescope resolution, the shadow of Sgr A* should be observable with 
very long baseline interferometry (VLBI) below $\sim 1$ mm; it should be
quite distinctive at $0.6$ mm.

During this past year, considerable progress has been made in modeling
the structure and physical attributes of the inner emitting region
surrounding Sgr A*, taking into account several crucial new data,
including mm/sub-mm polarization measurements (Aitken et al. 2000) and
{\it Chandra's} identification of an X-ray counterpart (Baganoff et
al.  2001).  In their analysis, Melia, Liu, \& Coker (2000, 2001)
concluded that the sub-mm bump is probably produced by thermal
synchrotron emission in a tight Keplerian flow within the inner $\sim
5\,r_S$.  A circularization such as this is expected on the basis of
hydrodynamical simulations which show that infalling gas possesses an
average specific angular momentum to settle into a rotating structure
below $\sim 40\,r_S$ (Coker \& Melia 1997).  Presumably the spectrum
of Sgr A* longward of $\sim 1-2$ mm is generated outside of the
Keplerian region, where the gas is making a transition from a
quasi-spherical infall into a circularized pattern. Within the
rotating plasma, self-Comptonization of the sub-mm bump radiation
produces an X-ray component that may account for the {\it Chandra}
source. 

Although this correlation between the radio and X-ray components in the 
spectrum of Sgr A* is not unique to this picture (see, e.g., Falcke \& 
Markoff 2000), the recent detection of significant linear 
polarization at mm and sub-mm wavelengths (if confirmed) seems to provide 
supporting evidence for the circularized flow when the flip of the
polarization vector is taken into account (Melia, Liu \& Coker 2000). 
The reason for this is that the optically thick emission is dominated by 
emitting elements on the near and far sides of the black hole, for which the 
Extraordinary wave (with the magnetic field pointing in the azimuthal
direction) has a polarization vector parallel to the reference 
axis (perpendicular to the Keplerian plane). In contrast, the dominant 
contribution in the thin region comes from the blue shifted emitter to the 
side of the black hole, where the Extraordinary wave has a polarization 
vector mostly perpendicular to this axis. This is why the position angle 
apparently shifts by about $90^\circ$ near the peak of the mm/sub-mm bump.

In this {\it Letter}, we present the results of new calculations that
update these findings in several very important ways, particularly
with regard to the upcoming VLBI imaging observations.
We have carried out fully self-consistent ray-tracing simulations with the
magnetized Keplerian flow geometry to incorporate the effects of
light-bending and area amplification into this observationally-motivated 
physical model.  One of our main goals is to produce 
{\it polarimetric} images motivated by the expected capability of mm and
sub-mm interferometry.

\section{Calculation of the Polarimetric Image}
Magnetic field dissipation suppresses the field intensity well below
its equipartition value in the quasi-spherical portion of the accreting
plasma (Kowalenko \& Melia 2000). However, once the gas circularizes
and settles into a Keplerian flow, a magnetohydrodynamic dynamo can
produce an enhanced (though still sub-equipartition) magnetic field,
dominated by its azimuthal component (Hawley, Gammie \& Balbus
1996).  Overcoming the rate of field destruction in the differentially
rotating portion of the inflow, the field reaches a saturated intensity
since the dynamo time scale is shorter than the dissipation time scale in
this region (Melia, Liu \& Coker 2001).  Our approach in this paper is to
consider a physical scenario that seems to account well for all the available
data, not only the mm/sub-mm spectrum and degree of polarization, but also
the position-angle flip, in order to predict the polarimetric map of Sgr A*.
Motivated by certain aspects of earlier hydrodynamical simulations and a
pseudo-Newtonian treatment of the inner Keplerian region, and combining
these with the phenomenology associated with the salient observational
constraints, we are led to adopt a model in which the gas circularizes
at small radii, winding the magnetic field within a relativistic Keplerian
disk.  We take the specific analysis of Melia, Liu, \& Coker (2000,2001)
to estimate the plausible structure of the gas distribution within
several Schwarzschild radii of the black hole, from which we infer the
emissivities required to produce the images.  We thus track both
the Extraordinary and Ordinary waves, which can be expressed, respectively, 
as\break 
\begin{eqnarray}\label{eq:eps}
\epsilon^e&=& {\sqrt{3} e^3\over 8\pi m_e c^2} B \sin{\theta^\prime} 
\int_0^\infty N(E)[F(x)+G(x)]\ dE\ , \label{com1} \\
\epsilon^o&=& {\sqrt{3} e^3\over 8\pi m_e c^2} B \sin{\theta^\prime} 
\int_0^\infty N(E)[F(x)-G(x)]\ dE\;, \label{com2}
\end{eqnarray}
where $N(E)$ is the electron distribution function at energy $E$, and
\begin{eqnarray}
\cos{\theta^\prime}= {\cos{\theta}-v_\phi/c\over 1-(v_\phi/c)\cos{\theta}}\; ,
&\qquad& x={4\pi\nu m_e^3c^5\over 3eB\sin{\theta^\prime}E^2}\; ,\\
F(x)= x\int_x^\infty K_{5/3}(z)\ dz\;,&\qquad&
G(x)= x\ K_{2/3}(x)\;.
\end{eqnarray}
$K_{5/3}$ and $K_{2/3}$ are the corresponding modified Bessel
functions (Pacholczyk 1970).  Here, $\nu$ is the frequency measured in
the co-moving frame, and $\theta$ is the angle between the local
magnetic field vector and the outward pointing ray.  We take a thermal
distribution of electrons to fix $N(E)$.  (Note that the temperature
in this region may reach $\sim 10^{11}$~K, for which inverse Compton
processes then produce an X-ray component.  We shall not
include this in our calculations since we are primarily interested in
mapping the polarization vector at mm/sub-mm wavelengths.)

In adopting the model of Melia, Liu, \& Coker (2001), we do not fix
the parameters a priori, but rather perform a $\chi^2$-minimization with
the goal of producing polarimetric images consistent with the spectral
and polarization data.
As we shall see, the simulation that produces a good fit to the
observed spectrum begins with a gas inflow rate of $3.6\times
10^{16}$~g~s$^{-1}$, at a circularization radius of 8~$r_S$.
The light-emitting flow extends down to
about 2~$r_S$, at which point the gas achieves a plunging orbit
into the black hole. Calculating the physical profiles as functions of
radius using the prescription in Melia, Liu \& Coker (2001), we find
that the temperature stays within a factor of a few times $10^{10}$~K
in the Keplerian region, while the particle density varies in the
range of about $2\times 10^7$ to $4\times 10^8$ cm$^{-3}$ and the magnetic field
increases from about 4 to 20 G.  With this model, we find that modestly
low inclination angles, around $30^\circ$ give a good fit to the observations,
although higher values are not strongly excluded. 

We emphasize that the model parameters adopted here provide a
self-consistent description of the flow in the neighborhood of the
supermassive black hole.  The inflow rate we infer is substantially
lower than the limits obtained using other arguments (e.g., Faraday
depolarization), considered within the context of alternative mechanisms
for producing the polarized emission (see, e.g., Agol 2000; Quataert
\& Gruzinov 2000). The viability of the model ultimately depends on
the uncertain conditions of the inflow (and, possibly, outflow) at larger
radii.  For example, the hydrodynamical simulations discussed earlier
are rather limited in scope, given that their modest resolution imposed
an inner boundary of $1,000r_S$ or more.  Thus, there is little guidance
as to the behavior of the gas at smaller radii, particularly with regard
to whether mass loss occurs or not.  Nonetheless, it seems implausible to
assume that the specific angular momentum accreted inward of this boundary
should increase.  At the same time, the heterogeneous nature of the Galactic Center
suggests that material arriving in the vicinity of the black will have
at least some residual angular momentum, so circularization at small radii
seems inevitable.  

The overall specific intensity $I_\nu^{e,o}$ observed at infinity is
an integration of the emissivity $\epsilon^{e,o}$ over the path length
along geodesics.  To obtain $I_\nu^{e,o}$ from the emitting accreting
disk, we make the simplifying assumption that the disk is
geometrically thin, and that its vertical structure is a
uniform slab (e.g., Melia, Liu \& Coker, 2001).
Otherwise, our calculations incorporate all relativistic effects
including frame dragging, gravitational redshift, light bending,
and Doppler boosting. We use a second-order geodesic solver similar to
that introduced by Bromley, Chen \& Miller (1997), and trace photon
trajectories from a pixel-array detector all the way through a thick
spherical shell bounded by the outer disk radius and a surface near
the event horizon. We pick up contributions to the
observed flux every time the trajectory intersects the disk
midplane. In this way, the primary disk image (from photons 
traveling directly to the observer without crossing the midplane after
leaving the disk) and higher-order images (multiple midplane
crossings) are formed.

The output of our ray-tracing code is a pixelized image, with specific
intensities at the detector calculated from the relativistic invariant
$I_\nu/\nu^3$. Emitter-frame frequencies come from the projection of
the emitted photon 4-momentum onto the 4-velocity of the emitter
(e.g., Cunningham 1975).  Similarly, local emission angles, such as
$\theta^{\prime}$ in Equation~(\ref{eq:eps}), 
follow from projection of the photon
4-momentum onto the spacelike components of a tetrad tied to the
emitter frame. We assume Keplerian flow except at
small radii where circular orbits are unstable. In this case the
emitters are on freefall trajectories as if perturbed from the minimum
stable orbit (3~$r_S$ for a black hole with zero angular momentum).
We calculate fluxes by summing over the pixels in the image, taking
into account the physical size of the detector array and the distance
to the disk.

The ray-tracing method adopted here is tuned for the synchrotron problem in
two novel ways. First, our code is capable of integrating
radiative transfer equations along photon trajectories. Here we make
only modest use of this feature by allowing for absorption of light from
high-order images after multiple disk crossings, taking into account the
redshift of photons as they travel from one region of the emitting gas
to another.  It will be more important to exercise the code's full
capabilities in future work when we treat detailed models of the disk's
vertical structure.

The second key feature of our code is the calculation of polarization
in a strongly curved spacetime.  It is straightforward to relate the
specific intensity of each polarization separately in the emitter and
detector frames from the relativistic invariant $I_\nu/\nu^3$, and
the degree of polarization is itself an invariant.  The position angle of
polarized light may also be calculated from a relativistic invariant
related to the parallel transport of a polarization vector along a
null ray (Conners \& Stark 1977; Laor, Netzer \& Piran 1991).  
Here we perform the parallel transport operation quite directly by
defining a reference vector at the detector and numerically propagating
it along with the null ray itself.  This allows us to consistently map 
a position angle from one frame to another, an essential feature for 
calculating radiative transfer for the Ordinary and Extraordinary waves 
that make multiple passes through the disk.

Aitken et al. (2000) provide specific constraints for the
magnetohydrodynamic disk model and our relativistic calculations in
terms of the total flux and polarization of Sgr~A*. The polarization
``flip'' in the spectrum near a wavelength of 1~mm is a particularly 
important constraint.  We use a nonlinear least-squares
search of the model parameter space, including spatial extent of the
accretion flow, the mass accretion rate, and other factors governing
the radial infall velocity, disk thickness and magnetic field strength
(see Melia, Liu \& Coker 2001 for details) in order to optimize the fit. 
Here we consider only a non-rotating black hole. The inclination angle 
$i$ is sampled sparsely, but sufficiently to demonstrate that low
values of $i$ provide a greater total flux, while high values of $i$
give a sharper polarization flip. As mentioned above we find that 
$i=30^\circ$ yields a good fit, although, larger angles are
not strongly excluded.

There are strong observational constraints on the linear polarization
from Sgr~A* at longer wavelengths as well.  Bower et al. (1999) 
have placed limits on the linear polarization at a variety of
frequencies below 100~GHz.  For our simulations, the 1\% upper bound
at 86~GHz is the most constraining; at this frequency, our simulation
predicts 0.7\% linear polarization. The predicted linear polarization
at longer wavelengths is $\ll$~0.1\%, consistent with the 
observationally-imposed limits.  Figure~1 shows a comparison between 
the simulation results and the data. 

\section{Discussion}
The polarimetric images demonstrate that future mm/sub-mm
interferometry can directly reveal material flowing near 
the horizon in Sgr~A*.  Here, the ``shadow'' of the black hole
takes the form of a dark cavity in a crescent. The bright crescent
shape itself, and its highly distinctive distribution of polarized
light, may be understood as a manifestation of the relativistic
beaming from material on the incoming side of the disk.  For optically
thin synchrotron emission, the specific intensity in the emitter frame
is highly anisotropic, and the beaming effect causes an alignment
between the observed photon trajectory and the direction of maximum
emission only on the incoming side. This effect is most apparent in
the images which show individual components of polarized light. For
example, in the absence of relativity the Extraordinary emission will
dominate the flux along the disk spin axis where the photon emission
and polarization position angles are at 90$^\circ$ to the azimuthal
magnetic field. However, the relativistic beaming causes the region of
maximum Extraordinary emission to shift toward the incoming side of
the disk.

Following Falcke, Melia \& Agol (2000) we now assess the possibility
that VLBI imaging may reveal the structure of the accretion flow, its
polarized emission, and the presence of the event horizon.  Figure~2 
shows images of the accretion flow at an ideal detector, as well as 
polarimetric images taking into account the blurring due to a finite 
VLBI resolution and scattering by the interstellar medium. The blurring 
is modeled by convolution with two Gaussians: the first is an approximate 
model of the scattering effects with an ellipsoidal filter whose major- 
and minor-axis FWHM values are 24.2~$\mu$as~$\times (\lambda/1.3\;\hbox{\rm 
mm})^2$ and 12.8~$\mu$as~$\times (\lambda/1.3\;\hbox{\rm mm})^2$, respectively,
for emission at wavelength $\lambda$ (Lo et al. 1998). For the purposes 
of creating test VLBI images, we chose the spin axis of the disk to lie 
along the minor axis of the scattering ellipsoid, and we have assumed a 
global interferometer with a $8,000$ km baseline (Krichbaum 1996).
The second is a spherically symmetric filter with a FWHM of 
33.5~$\mu$as~$\times (\lambda/1.3\;\hbox{\rm mm})$ to account for the
resolution effects of an ideal interferometer.  These images demonstrate
clearly the viability of conducting polarimetric imaging of the
black hole at the Galactic center with upcoming VLBI techniques.

{\bf Acknowledgments} We thank Vladimir Pariev for his suggestions
on the polarization calculations in curved spacetime.  This research was
partially supported by NASA under grants NAG5-8239, NAG5-9205, and
NAG5-8277, and has made use of NASA's Astrophysics Data System
Abstract Service.  FM is very grateful to the University of Melbourne
for a Miegunyah Fellowship. BCB acknowledges computing support from NASA/JPL Supercomputing.

%
%
\newpage

\begin{figure}[thb]\label{fig:spec}
\epsscale{0.8}
\centerline{\plotone{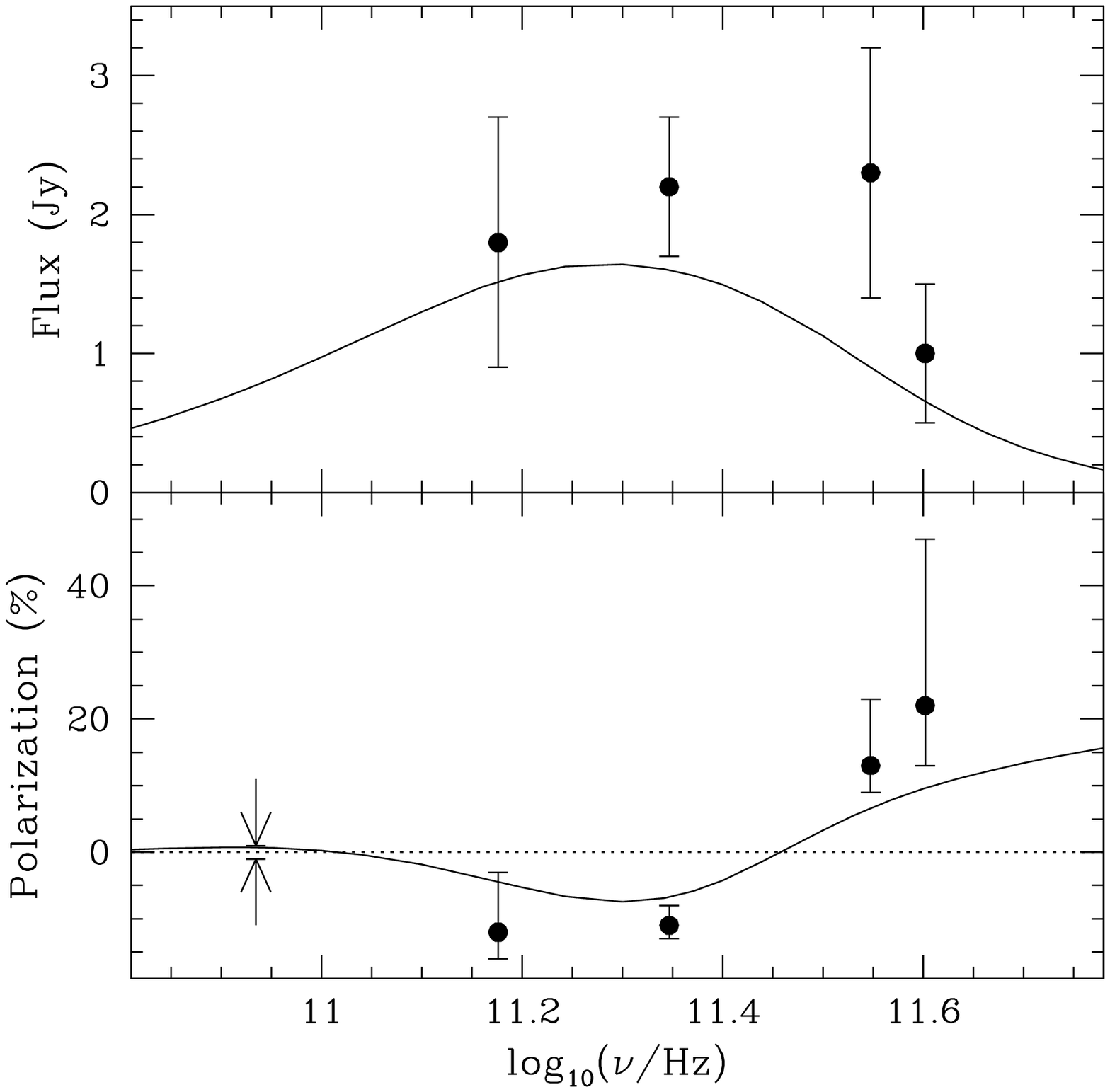}}
\caption{Spectra showing total flux density and polarization from Sgr A*.
The high-frequency data points are from Aitken et al.~(2000) and the curves
are from the magnetohydrodynamic model discussed in the text.  The
limit at 84 GHz is from Bower et al. (1999).  At even lower
frequencies, the best fit polarization is consistent with $0$.  In this
figure, `negative' polarization corresponds to the polarization vector
being aligned with the spin axis of the black hole, whereas positive
is for a perpendicular configuration.  The polarization crosses $0$
when this vector flips by $90^\circ$.
}
\end{figure}

\clearpage
\begin{figure}[thb]\label{fig:maps}
\centerline{\plotone{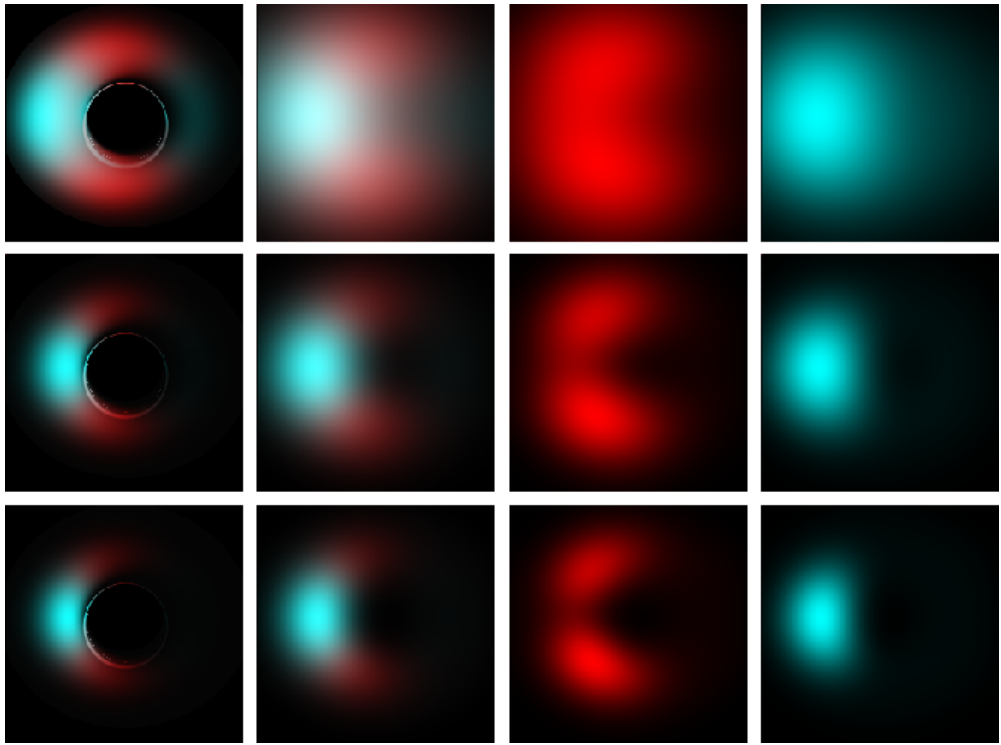}}
\caption{Polarization maps at three wavelengths near the
peak of the mm to sub-mm emission from Sgr~A*. The top row shows
emission at 1.5~mm, the middle row is at 1~mm, and the
bottom row corresponds to 0.67~mm. The images in each row
show the raw ray-tracing output (first column on the left), and
an image blurred to account for finite VLBI resolution and
interstellar scattering (second column). The two rightmost
columns give the vertical and horizontal components of the polarized emission.
Throughout, red pixels designate vertically polarized light, and
cyan corresponds to horizontal polarization. The pixel brightness in 
all images scales linearly with flux. }
\end{figure}

\end{document}